\documentclass[12pt]{article}
\usepackage{graphicx}
\usepackage{amssymb}
\def\hybrid{\topmargin 0pt      \oddsidemargin 0pt
        \headheight 0pt \headsep 0pt
        \voffset=-0.5cm
        \textwidth 6.25in       % A4 paper
        \textheight 9.5in       % A4 paper
        \marginparwidth 0.0in
        \parskip 5pt plus 1pt   \jot = 1.5ex}
\catcode`\@=11
\def\marginnote#1{}

\newcount\hour
\newcount\minute
\newtoks\amorpm
\hour=\time\divide\hour by60 \minute=\time{\multiply\hour by60 \global\advance\minute by-\hour}
\edef\standardtime{{\ifnum\hour<12 \global\amorpm={am}%
        \else\global\amorpm={pm}\advance\hour by-12 \fi
        \ifnum\hour=0 \hour=12 \fi
        \number\hour:\ifnum\minute<10 0\fi\number\minute\the\amorpm}}
\edef\militarytime{\number\hour:\ifnum\minute<10 0\fi\number\minute}

\def\draftlabel#1{{\@bsphack\if@filesw {\let\thepage\relax
   \xdef\@gtempa{\write\@auxout{\string
      \newlabel{#1}{{\@currentlabel}{\thepage}}}}}\@gtempa
   \if@nobreak \ifvmode\nobreak\fi\fi\fi\@esphack}
        \gdef\@eqnlabel{#1}}
\def\@eqnlabel{}
\def\@vacuum{}
\def\draftmarginnote#1{\marginpar{\raggedright\scriptsize\tt#1}}
\def\draftlabel#1{{\@bsphack\if@filesw {\let\thepage\relax
   \xdef\@gtempa{\write\@auxout{\string
      \newlabel{#1}{{\@currentlabel}{\thepage}}}}}\@gtempa
   \if@nobreak \ifvmode\nobreak\fi\fi\fi\@esphack}
        \gdef\@eqnlabel{#1}}
\def\@eqnlabel{}
\def\@vacuum{}
\def\draftmarginnote#1{\marginpar{\raggedright\scriptsize\tt#1}}

\def\draft{\oddsidemargin -.5truein
        \def\@oddfoot{\sl preliminary draft \hfil
        \rm\thepage\hfil\sl\today\quad\militarytime}
        \let\@evenfoot\@oddfoot \overfullrule 3pt
        \let\label=\draftlabel
        \let\marginnote=\draftmarginnote
   \def\@eqnnum{(\theequation)\rlap{\kern\marginparsep\tt\@eqnlabel}%
\global\let\@eqnlabel\@vacuum}  }

\def\numberbysection{\@addtoreset{equation}{section}
        \def\theequation{\thesection.\arabic{equation}}}

\def\underline#1{\relax\ifmmode\@@underline#1\else
        $\@@underline{\hbox{#1}}$\relax\fi}

\def\titlepage{\@restonecolfalse\if@twocolumn\@restonecoltrue\onecolumn
     \else \newpage \fi \thispagestyle{empty}\c@page\z@
        \def\thefootnote{\fnsymbol{footnote}} }

\def\endtitlepage{\if@restonecol\twocolumn \else  \fi
        \def\thefootnote{\arabic{footnote}}
        \setcounter{footnote}{0}}  %\c@footnote\z@ }
\numberbysection \hybrid

\newcommand{\tr}{{\rm tr}}

\newcommand{\ti}[1]{\tilde{#1}}

\def\beq{\begin{equation}}
\def\eeq{\end{equation}}

\def\asl2{{\rm sl}(2, {\mathbb C})}
\def\GL2{{\rm GL}(2, {\mathbb C})}
\def\SL2{{\rm SL}(2, {\mathbb C})}

\begin{document}

\begin{titlepage}
\setcounter{page}{1}

\title{%On
Spectral Duality in Integrable Systems\\ from AGT Conjecture\\ $\ $}

\author{
A. Mironov
\thanks{Theory Department, Lebedev Physics Institute and ITEP, Moscow, Russia.
$\ \ \ \ \ \ \ \ \ \ \ \ \ \ \ $ $\ \ \ \ \ \ \ \ \ $ E-mail: mironov@itep.ru; mironov@lpi.ru} \and A. Morozov
\thanks{ITEP, Moscow, Russia, E-mail: morozov@itep.ru}\and
Y. Zenkevich
\thanks{Physical Department, Moscow State University, Institute for Nuclear
Research of the Russian Academy of Sciences and ITEP, Moscow, Russia. E-mail: yegor.zenkevich@gmail.com}\and A. Zotov
\thanks{ITEP, Moscow, Russia. E-mail: zotov@itep.ru}
}

\date{}

\maketitle

\vspace{-7cm} \centerline{ \hfill FIAN/TD-03/1} \centerline{\hfill ITEP-TH-13/12} \vspace{7cm}

\begin{abstract}
We describe relationships between integrable systems with $N$ degrees of freedom arising from the AGT conjecture. Namely, we
prove the equivalence (spectral duality) between the $N$-cite Heisenberg spin chain and a reduced  ${\rm gl}_N$ Gaudin model
both at classical and quantum level. The former one appears on the gauge theory side of the AGT relation in the
Nekrasov-Shatashvili (and further the Seiberg-Witten) limit while the latter one is natural on the CFT side. At the classical
level, the duality transformation relates the Seiberg-Witten differentials and spectral curves via a bispectral involution.
The quantum duality extends this to the equivalence of the corresponding Baxter-Schr\"odinger equations (quantum spectral
curves). This equivalence generalizes both the spectral self-duality between the $2\times 2$ and $N\times N$ representations
of the Toda chain and the famous AHH duality.
%\cite{AHH}.
%rank/number of points
\end{abstract}

%\vfill\eject %\tableofcontents

%\vfill

\end{titlepage}

\small{

\section{Introduction}
\label{sec:introduction}

In this paper we study the AGT correspondence \cite{AGT} at the level of integrable systems \cite{SWint,NS,BS}\footnote{See
also
\cite{AGTN},\cite{SO},\cite{AGTmamo},\cite{BSmore},\cite{MiTa},\cite{Yamada},\cite{5},\cite{Mironov:2010qe},\cite{Tai},\cite{Tai2}.}.
More exactly, we deal with the AGT inspired models which emerge in the limiting case. The full AGT correspondence associates
the conformal block of the Virasoro or $W$-algebra in two-dimensional conformal field theory with the LMNS integral
\cite{LMNS} (Nekrasov functions \cite{Nekr})) describing the two-parametric deformation of Seiberg-Witten theory by
$\Omega$-background. The classical integrable systems emerge when both deformation parameters are brought to zero, while when
only one of the parameters going to zero (the Nekrasov-Shatashvili limit \cite{NS}) the integrable system gets quantized. We
shall study here only the correspondence between AGT inspired integrable systems in these two limiting cases.

It is important that the two sides of the AGT correspond to {\it a priori} different types of integrable models which should
actually coincide due to AGT. This leads to non-trivial predictions of equivalence of different models and also illuminates
what the equivalence exactly means. Here we consider the simplest example of this kind: the equivalence of the four-point
conformal block and the prepotential in the $SU(N)$ SUSY theory with vanishing $\beta$-function. On the gauge theory side the
(classical) integrable system is known \cite{SWsc} to be the Heisenberg chain \cite{Heisen} which is described by the
spectral curve $\Gamma^{\hbox{\tiny{Heisen}}}(w,x):\det(w-T(x))=0$ with ${\rm GL}_2$-valued $N$-site transfer-matrix $T(x)$
and Seiberg-Witten \cite{SW} (SW) differential $\hbox{d}S^{\hbox{\tiny{Heisen}}}(w,x)=x\frac{\hbox{d}w}{w}$. On the CFT side
the corresponding integrable system was argued to be some special reduced Gaudin model \cite{Gaudin1} defined by  its
spectral curve $\Gamma^{\hbox{\tiny{Gaudin}}}(y,z):\det(y-L(z))=0$ with ${\rm gl}_N$-valued Lax matrix $L(z)$ and the SW
differential $\hbox{d}S^{\hbox{\tiny{Gaudin}}}(y,z)=y\hbox{d}z$. The original argument \cite{AGT} dealt with the $SU(2)$ case
and implied that on the conformal side of the AGT correspondence the counterpart of the SW differential is played by the
average of the energy-momentum tensor, and this latter shows up a pole behaviour which is rather associated with the Gaudin
model. This argument was refined later by associating the SW differential with an insertion of the surface operator
\cite{SO,MiTa,5} or with the matrix model resolvent \cite{AGTmamo}.

If considering the case of higher rank group $SU(N)$, which on the gauge theory side is associated with the Heisenberg chain
(on $N$ sites), one has to take into account that on the conformal side the AGT conjecture in this case deal with a
four-point conformal block of the $W_N$-algebra \cite{AGTN}, however not an arbitrary one but that restricted with special
conditions imposed onto two of the four external operators (states) of the block. This means that there are two arbitrary
operators parameterized by $N-1$ parameters each and two other operators parameterized by only one parameter each. In
integrable terms this means that one should expect for the associated integrable system, the reduced Gaudin model that it is
described by two coadjoint orbits of the maximal dimensions inserted in two points, and by two coadjoint orbits of the
minimal dimensions inserted in two other points. As we shall see, this is, indeed, the case.

In this letter we show that the change of variables $z=w$, $\lambda=x/w$ relates the curves and SW differentials of the two
integrable systems under discussion (the Heisenberg spin chain and the reduced Gaudin model). It means that with this change
of variables the following relations hold true:
 \begin{equation}\label{q51007}
  \begin{array}{c}
  \Gamma^{\hbox{\tiny{Gaudin}}}(y,z)=\Gamma^{\hbox{\tiny{Heisen}}}(z,zy)\,,
  \\
  \
  \\

\hbox{d}S^{\hbox{\tiny{Gaudin}}}(y,z)=\hbox{d}S^{\hbox{\tiny{Heisen}}}(z,zy)\,.
  \end{array}
 \end{equation}
This type of relations between spectral curves appeared in \cite{AHH}\footnote{We call the duality between Gaudin models
described in \cite{AHH} as AHH duality. See the comment in the end of the paper.}. Following \cite{Harnad2} we call it {\em
(classical) spectral duality}. The duality transformation acts by {\em bispectral involution} \cite{W1} which interchanges
the roles of the eigenvalue-variable and spectral parameter.

A well-known simpler example is the periodic Toda chain. It can be described by both the ${\rm gl}(N)$-valued Lax matrix:
 \begin{equation}\label{qq100} { L}^{Toda}_{N\times N}(z) =
\left(\begin{array}{ccccc}
p_1 & e^{{1\over 2}(q_2-q_1)} & 0 & & ze^{{1\over 2}(q_1-q_{N})}\\
e^{{1\over 2}(q_2-q_1)} & p_2 & e^{{1\over 2}(q_3 - q_2)} & \ldots & 0\\
0 & e^{{1\over 2}(q_3-q_2)} & p_3 & & 0 \\
 & & \ldots & & \\
\frac{1}{z}e^{{1\over 2}(q_1-q_{N})} & 0 & 0 & & p_{N}
\end{array} \right)
 \end{equation}
and the ${\rm GL}(2)$-valued  transfer-matrix:
 \begin{equation}\label{qq1010}
T^{Toda}_{2\times 2}(\lambda)=L_N(\lambda)...L_1(\lambda),\ \ \
L_i(\lambda) = \left(\begin{array}{cc} \lambda -p_i & e^{q_i} \\
-e^{-q_i} & 0
\end{array}\right), \ \ \ \ \ i = 1,\dots ,N
 \end{equation}
The spectral curves defined by these representations are related by the bispectral involution, i.e.
 \begin{equation}\label{qq102}
\det(\lambda-{ L}^{Toda}_{N\times N}(z))=0\ \ \ \hbox{and}\ \ \ \ \det(z-T^{Toda}_{2\times 2}(\lambda))=0
 \end{equation}
coincide. The SW differential is the same in both cases $\hbox{d}S=\lambda\frac{\hbox{d}z}{z}$. Therefore, {\em the periodic
Toda chain is a self-dual model} \cite{GGM}.

The quantum version of  the duality  appears from the exact quasi-classical quantization of the spectral curves. Considering
the SW differential as a symplectic 1-form \cite{Krich} on ${\mathbb C}^2$-plane $(y,z)$ yields a pair of canonical variables
$(p(y,z),q(z))$ which brings the SW differential to $\hbox{d}S(y,z)=p\hbox{d}q$. Then there is a natural quantization of the
spectral curve defined by the rule $(p,q)\rightarrow (\hbar\partial_q, q)$. For the above mentioned models one has:
 \begin{equation}\label{qq10205}
 \begin{array}{c}
{\hat\Gamma}^{\hbox{\tiny{Heisen}}}(z,\hbar z\partial_z)\Psi^{\hbox{\tiny{Heisen}}}(z)=0\,,
 \end{array}
 \end{equation}
 \begin{equation}\label{qq10208}
 \begin{array}{c}
\hat\Gamma^{\hbox{\tiny{Gaudin}}}(\hbar
\partial_z,z)\Psi^{\hbox{\tiny{Gaudin}}}(z)=0
 \end{array}
 \end{equation}
with some choice of ordering. The wave functions can be written in terms of the quantum deformation of the SW differential on
the spectral curve, i.e. $\Psi(z)=\exp\left(-\frac{1}{\hbar}\int^q\hbox{d}S(\hbar)\right)$, where
$\hbox{d}S(\hbar)=p(q,\hbar)\hbox{d}q$ and $p(q,0)=p(q)|_\Gamma$. The monodromies of the wave function around $A$- and $B$-
cycles of $\Gamma$ are given by the quantum deformed action type variables \cite{BS}:
 \begin{equation}\label{qq10206}
 \begin{array}{l}
\Psi(z+A_i)=\exp\left(-\frac{1}{\hbar}a_i^\hbar\right)\Psi(z),\ \ a_i^\hbar=\oint\limits_{A_i}\hbox{d}S(\hbar)\,,
\\
  \
  \\
\Psi(z+B_i)=\exp\left(-\frac{1}{\hbar}\frac{\partial \mathcal{F}_{\hbox{\tiny{NS}}}}{\partial a_i^\hbar}\right)\Psi(z),\ \
\frac{\partial \mathcal{F}_{\hbox{\tiny{NS}}}}{\partial a_i^\hbar}=\oint\limits_{B_i}\hbox{d}S(\hbar)\,,
 \end{array}
 \end{equation}
where $\mathcal{F}_{\hbox{\tiny{NS}}}$ is the Nekrasov-Shatashvili limit \cite{NS} of the LMNS integral \cite{LMNS}.

The AGT conjecture predicts the following relations ({\em quantum spectral duality}):
 \begin{equation}\label{qq900}
 \begin{array}{l}
a_i^\hbar(\Psi^{\hbox{\tiny{Heisen}}})=a_i^\hbar(\Psi^{\hbox{\tiny{Gaudin}}})\,,
\\
  \
  \\
\frac{\partial \mathcal{F}_{\hbox{\tiny{NS}}}}{\partial a_i^\hbar}(\Psi^{\hbox{\tiny{Heisen}}})=\frac{\partial
\mathcal{F}_{\hbox{\tiny{NS}}}}{\partial a_i^\hbar}(\Psi^{\hbox{\tiny{Gaudin}}})\,.
 \end{array}
 \end{equation}

In this paper we deal with the known quantum equation (\ref{qq10205}) for the XXX chain - the Baxter equation\footnote{It
arises as an equation for the Baxter Q-operator eigenvalues in the Quantum Inverse Scattering Method. Originally, it was
written in difference (Fourier-dual) form.} \cite{Baxter}:
 \begin{equation}
  \label{q70}
  \left( \tr \, T(\hbar z \partial_z) - \frac{z}{1+q} {K}_{+}(\hbar
  z \partial_z) - \frac{q}{(1+q)z} {K}_{-}(\hbar
  z \partial_z) \right)\Psi^{\hbox{\tiny{Heisen}}}(z)=0\,.
 \end{equation}
We verify that (\ref{q70}) can be re-written as the quantum spectral curve of the Gaudin model (\ref{qq10208}). In this way
we arrive to the quantum version of duality:
 \begin{equation}\label{qq10207}
 \begin{array}{|c|}
  \hline\\
\Psi^{\hbox{\tiny{Heisen}}}(z)=\Psi^{\hbox{\tiny{Gaudin}}}(z)\\ \ \\ \hline
  \end{array}
 \end{equation}

In the next section we briefly describe the models and formulate the spectral duality. Some comments are given at the end.
Most of details will be given in \cite{MMRZZ}. In that extended version we also plan to describe the Poisson map between
models.

{\footnotesize{

\paragraph{Acknowledgments} The authors are grateful to A.Gorsky,  A.Zabrodin
and A.Zhedanov for useful comments and remarks. The work was partially supported by the Federal Agency for Science and
Innovations of Russian Federation under contract 14.740.11.0347 (A.Z., Y.Z.), by NSh-3349.2012.2, by RFBR grants 10-02-00509
(A.Mir.), 10-02-00499 (A.Mor., Y.Z.), 12-01-00482 (A.Z.) and by joint grants 11-02-90453-Ukr, 12-02-91000-ANF,
12-02-92108-Yaf-a, 11-01-92612-Royal Society. The work of A.Zotov was also supported in part by the Russian President fund
MK-1646.2011.1.

}}

\section{Heisenberg Chain - Gaudin Model Duality}
\label{sec:xxx-heisenberg-chain}

\noindent{\bf 1. $N$-site ${\rm GL}_2$ Heisenberg (XXX) chain}. It is classically defined by its spectral curve
 \begin{equation}\label{q14}
  \Gamma^{\hbox{\tiny{Heisen}}}(w,x):\ \
  \tr T(x)-\frac{1}{1\!+\!q}wK^+(x)-\frac{q}{1\!+\!q}w^{-1}K^-(x)=0,\
K^\pm(x)=\prod\limits_{i=1}^N(x-m_i^\pm)
 \end{equation}
and SW differential
 \begin{equation}\label{q3011}
  \hbox{d}S^{\hbox{\tiny{Heisen}}}(w,x)=x\frac{\hbox{d}w}{w}\,.
 \end{equation}
$T(x)$ in (\ref{q14}) is ${\rm GL}_2$-valued transfer-matrix:
 \begin{equation}\label{q2}
  \begin{array}{c}
  T(x)= V L_N(x) \ldots L_1(x),\ \ L_i(x)\!=\!x\!-\!x_i\!+\!S^i,\  i = 1
\ldots N,\ \

V={{\left(\begin{array}{cc}{1}&{-\frac{q}{(1+q)^2}}\\{1}&{0}\end{array}\right)}}\,,\\
  S^i\in  {\rm sl}_2:\ \hbox{Spec}(S^i)=(K_i,-K_i),\ \ m_i^\pm=x_i\pm K_i\,.
  \end{array}
 \end{equation}
Function
 \begin{equation}\label{q3012}
  \tr T(x)=x^N + \sum\limits_{i=1}^N x^{i-1}H^{\hbox{\tiny{Heisen}}}_i
 \end{equation}
provides commuting integrals of motion.

\noindent{\bf 2. Special (reduced) ${\rm gl}_N$ Gaudin model on ${\mathbb{CP}}^1\backslash\{0,1,q,\infty\}$}. It is described
by the spectral curve
 \begin{equation}\label{q3013}
  \Gamma^{\hbox{\tiny{Gaudin}}}(y,z):\ \ \det (y-L(z))=0,\ \ \
L(z)=\frac{A^0}{z}+\frac{A^1}{z-1}+\frac{A^q}{z-q}\in
  {\rm gl}_N
 \end{equation}
with additional conditions including the reduction constraints\footnote{One should also fix the action of the Cartan
subgroup. We do not discuss it here since it does not effect the curve.}
 \begin{equation}\label{q31}
 \begin{array}{c}
  A^0+A^1+A^q+A^\infty=0\,,\\
  A^\infty\equiv\Upsilon=\hbox{diag}(\upsilon_1,...,\upsilon_N),\ \
\hbox{Spec}(A^0)=(\mu_1,...,\mu_N)\,,\\
  A^1=\xi^1\times\eta^1,\ \ A^q=\xi^q\times \eta^q\,,
  \end{array}
 \end{equation}
i.e. $A^1$ and $A^q$ are ${\rm gl}_N$ matrices of rank 1 (this type of configuration was already discussed
\cite{Tsuda,Yamada}). Using specification (\ref{q31}) the spectral curve can be find explicitly:
 \begin{equation}\label{q40}
  \begin{array}{c} \left(

\eta^1(zy\!+\!\Upsilon)^{\!-1}\xi^1\!+\!q\eta^q(zy\!+\!\Upsilon)^{\!-1}\xi^q \!+\!q\!+\!1\right)\!\prod\limits_{i=1}^N
      (zy\!+\!\upsilon_i)
      =z\!\prod\limits_{i=1}^N(zy\!+\!\upsilon_i)\!
      +z^{\!-1}\!q\!\prod\limits_{i=1}^N(zy\!-\!\mu_i)
  \end{array}
 \end{equation}
or
 \begin{equation}\label{q4003}
  \begin{array}{c}
\prod\limits_{i=1}^N(zy+\upsilon_i)+ \sum\limits_{k=1}^N\frac{\eta^1_k\xi^1_k+q\eta^q_k\xi^q_k}{q+1}\prod\limits_{i\neq
k}^N(zy+\upsilon_i) = \frac{z}{q+1}\prod\limits_{i=1}^N(zy+\upsilon_i)
      \!+\!z^{-1}\frac{q}{q+1}\prod\limits_{i=1}^N(zy-\mu_i)\,.
  \end{array}
 \end{equation}

The SW differential is
 \begin{equation}\label{q2002}
  \hbox{d}S^{\hbox{\tiny{Gaudin}}}(y,z)=y\hbox{d}z\,.
\end{equation}
{\bf The classical spectral duality.} First, notice that the both models (as classical mechanical systems) describe dynamics
of $N-1$ degrees of freedom and depend on $2N+1$ parameters.

Indeed, the dynamical variables of the off-shell Gaudin model (\ref{q31}) are $A^{0,1,q,\infty}$. Fixing the Casimir
functions restricts $A^{0,1,q,\infty}$ to the coadjoint orbits of maximum dimensions ($N^2-N$) at $z=0,\ \infty$ and of
minimal dimensions ($2N-2$) at $z=1,\ q$. Then the reduction by the coadjoint action of ${\rm GL}_N$ gives the following
dimension of the phase space:
 \begin{equation}\label{q2007}
2(N^2-N)+2(2N-2)-2(N^2-1)=2(N-1).
\end{equation}
The number of parameters is $2N+3:\ \{\upsilon_1,...,\upsilon_N,\mu_1,...,\mu_N,\tr A^1,\tr A^q,q\}$. Two of them, ($\tr A^0,
\tr A^\infty$) can be eliminated from the spectral curve by the shift of $y$. Therefore, the number of independent parameters
is $2N+1$.

For the Heisenberg chain, one initially has $N$ ${\rm sl}_2$-valued variables $S^i$ with the Casimir functions fixed at each
site: $\frac{1}{2}\tr\left(S^i\right)^2=K_i^2$. The reduction by $\hbox{Stab}(V(q))\cong \hbox{Cartan}({\rm GL}_2)$ fixes two
independent variables. Therefore, for the dimension of the phase space one has
 \begin{equation}\label{q2008}
3N-N-2=2(N-1)
\end{equation}
and there are $2N+1$ parameters $\{x_1,...,x_N,K_1,...,K_N,q\}$.

The duality between models is described by the following

{\bf{Theorem.\ }}{\em The N-site  ${\rm GL}_2$ Heisenberg XXX chain defined by (\ref{q14})-(\ref{q3012}) and  the ${\rm
gl}_N$ Gaudin model (\ref{q3013})-(\ref{q2002}) are spectrally dual at the classical level
 \begin{equation}\label{q51005}
  \begin{array}{|c|}
  \hline\\
  \Gamma^{\hbox{\tiny{Gaudin}}}(y,z)=\Gamma^{\hbox{\tiny{Heisen}}}(w,x)
  \\
  \
  \\
  \hbox{d}S^{\hbox{\tiny{Gaudin}}}(y,z)=\hbox{d}S^{\hbox{\tiny{Heisen}}}(w,x)
  \\
  \
  \\
  \hline
  \end{array}
 \end{equation}
with the following change of variables
 \begin{equation}\label{q51006}
    \begin{array}{c} z=w,\ \ y=\frac{x}{w}\,,
    \end{array}
 \end{equation}
identification of parameters
 \begin{equation}\label{q57006}
    \begin{array}{c}
m^+_i=-\upsilon_i,\ \
        m^-_i=\mu_i,\ \ 1\leq
        i\leq N\,,
    \end{array}
 \end{equation}
and relation between generating functions of the Hamiltonians:
  \begin{equation}\label{q53}
    \begin{array}{c}
      \tr T^{\hbox{\tiny{Heisen}}}(y)= \det(y\!+\!\Upsilon)\left(1\!+\!

\frac{1}{1+q}\eta^1(y\!+\!\Upsilon)^{\!-1}\xi^1+\frac{q}{1+q}\eta^q(y\!+\!\Upsilon)^{\!-1}\xi^q\right)\,.
    \end{array}
  \end{equation}
}
\noindent The statement follows from   the  comparison of (\ref{q14}) and (\ref{q40}). In particular,
 \begin{equation}\label{q57007}
H^{\hbox{\tiny{Heisen}}}_N=\frac{1}{1+q}\tr A^1+\frac{q}{1+q}\tr
        A^q+\sum\limits_{k=1}^N\upsilon_k\,.
 \end{equation}

{\bf The quantum spectral duality.} The quantization of the XXX chain spectral curve (\ref{q40}) with the SW differential
(\ref{q3011}) means that $x$ should be simply replaced by $\hbar w\partial_w$. Then one gets the Baxter equation:
 \begin{equation}
  \label{qq912}
  \left( \tr \, T(\hbar w \partial_w) - \frac{w}{1+q} {K}_{+}(\hbar
  w \partial_w) - \frac{q}{(1+q)w} {K}_{-}(\hbar
  w \partial_w) \right)\Psi^{\hbox{\tiny{Heisen}}}(w)=0\,.
 \end{equation}
Equivalently, for the Gaudin spectral curve (\ref{q4003}) the quantization is given by the replacement
$y\rightarrow\hbar\partial_z$:
 \begin{equation}\label{qq914}
  \begin{array}{l}
\left(\prod\limits_{i=1}^N(z\hbar\partial_z+\upsilon_i)+
\sum\limits_{k=1}^N\frac{\eta^1_k\xi^1_k+q\eta^q_k\xi^q_k}{q+1}\prod\limits_{i\neq
k}^N(z\hbar\partial_z+\upsilon_i) -\right.\\
 \left.-\frac{z}{q+1}\prod\limits_{i=1}^N(z\hbar\partial_z+\upsilon_i)

-z^{-1}\frac{q}{q+1}\prod\limits_{i=1}^N(z\hbar\partial_z-\mu_i)\right)\Psi^{\hbox{\tiny{Gaudin}}}(z)=0\,.
  \end{array}
 \end{equation}
Obviously, the differential operators in the brackets of (\ref{qq912}) and (\ref{qq914}) can be identified in the same way as
the classical spectral curves did.

\section{Comments}
\label{comments}

\begin{itemize}

\item

{\em AHH duality.} In \cite{AHH} (see also \cite{BPMY}) the authors considered the Gaudin model with $M$ marked  points and
the Lax matrix defined as follows:
 \begin{equation}\label{q71}
L^G_{AHH}(z)=Y+\sum\limits_{c=1}^M \frac{A^c}{z-z_c}\,,\ \ Y= \hbox{diag }(y_1,...,y_N)\,,\ \  A^c\in {\rm gl}_N\,.
 \end{equation}
The later differs from ours. The difference is significant since $Y\neq 0$ leads to the second order pole at $\infty$ for
$L^G_{AHH}(z)\hbox{d}z$. The phase space is also different. It is a direct product of the coadjoint orbits (equipped with a
natural Poisson-Lie structure) factorized by  the stabilizer of $Y$: ${\mathcal O}^1\times\dots\times{\mathcal
O}^M//\hbox{Stab}(Y)$.

In the case when all $A^c$ are of rank 1 the dual Lax matrix is the ${\rm gl}_M$-valued function with $\ti Y=\hbox{diag
}(z_1,...,z_M)$ and $N$ marked  points at $y_1,...,y_N$:
 \begin{equation}\label{q7101}
{\ti{L}}^G_{AHH}(z)=\ti Y+\sum\limits_{c=1}^N \frac{\ti A^c}{z-y_c}\,,\ \ \ti Y= \hbox{diag }(z_1,...,z_M)\,, \ \ \ti A^c\in
{\rm gl}_M\,.
 \end{equation}
The duality implies the following relation between the spectral curves:
 \begin{equation}\label{q7102}
\det(\ti Y-z)\det(L^G_{AHH}(z)-\lambda)=\det(Y-\lambda)\det(\ti L^G_{AHH}(\lambda)-z)\,.
  \end{equation}
The dimensions of the phase spaces of both models equal $2(N-1)(M-1)$ and the number of parameters is $N+M-1$.

\item

Sometimes ${\rm sl}_N$ description of the Gaudin model is more convenient than the ${\rm gl}_N$ one. The transformation of
the spectral curve from ${\rm gl}_N$ to ${\rm sl}_N$  is given by the simple shift:
 \begin{equation}\label{q57008}
 y\rightarrow y'=y-\frac{1}{N}\tr
L(z)=y-\frac{1}{zN}\left(-\tr\Upsilon+\frac{\tr A^1}{z-1}+q\frac{\tr A^q}{z-q}\right)\,.
 \end{equation}
In this case the change of variables (\ref{q51006}) is modified\footnote{In this form the change of variables was found in
\cite{Mironov:2010qe} for ${\rm sl}_2$ case.}:
 \begin{equation}\label{qq902}
    \begin{array}{c} z=w,\ \ y'=\frac{x-R(z)}{w},\ \
    R(z)=\frac{1}{N}\left(-\tr\Upsilon+\frac{\tr A^1}{z-1}+q\frac{\tr
A^q}{z-q}\right)\,.
    \end{array}
 \end{equation}
The equality of the wave functions (\ref{qq10207}) acquires the predictable multiple:
 \begin{equation}\label{qq922}
    \begin{array}{c}
\Psi^{\hbox{\tiny{Heisen}}}(z)=\Psi^{\hbox{\tiny{Gaudin}}}(z)e^{\frac{1}{N\hbar}\int^z b_h(z)\hbox{d}z}\,,\\ \ \\b_{\hbar}
(z) = \frac{(1+q)}{(z-1)(z-q)} \left( {H^{\hbox{\tiny{Heisen}}}_N} +
    \frac{z \sum_{k=1}^N m_k^{+}}{1+q} + \frac{q \sum_{k=1}^{N}
      m_k^{-}}{(1+q) z } \right) - \hbar \frac{N(N-1)}{2z}\,.
    \end{array}
 \end{equation}

\item

It should be mentioned that we do not impose any boundary conditions which provide a valuable quantum problem, i.e. we do not
specify wave functions explicitly. To compare the quantum problems one needs a construction of the Poisson (and then quantum)
map between the phase spaces (Hilbert spaces) of the two models. We are going to describe the Poisson map elsewhere
\cite{MMRZZ}.

Alternatively, one can specify the spaces of solutions initially and then verify their identification through the duality
transformation. This is the recipe of \cite{MTV} where the authors considered very close problem in terms of the Bethe
vectors. The precise connection between the two approaches deserves further elucidation. We will comment on it in
\cite{MMRZZ}.

\item Besides the approach proposed here, a quantization of the Gaudin model
is known from \cite{FFR} and \cite{Talalaev}. We hope to shed light on relations between the quantizations in further
publications.

\item At last, let us mention possible generalizations of the correspondence
proposed in this letter. First of all, one can naturally consider multi-point conformal blocks. This provides one with the
multi-point Gaudin model. At the same time, the AGT predicts in this case on the other side of the correspondence the theory
with gauge group being a product of a few gauge factors. This latter is naturally embedded into the spin magnets with higher
rank group \cite{GGM}. Thus, one expects a correspondence between $GL(p)$-magnets and multi-point Gaudin models.

Another interesting generalization is induced by the five-dimensional AGT \cite{5dAGT} which implies a correspondence between
the XXZ magnets (see \cite{SW5d}) and a Gaudin-like model with relativistic (difference) dynamics. This latter would emerge,
since on the conformal side one deal in this case with the q-Virasoro conformal block which implies a difference
Schr\"odinger equation for the block with insertion of the degenerate field. An extension to six dimensions (elliptic
extension of the differential operator in the Schr\"odinger equation versus XYZ magnet) is also extremely interesting to
construct.

As is well known, the ${\rm sl}_2$ reduced Gaudin model with the configuration discussed above can be written in different
elliptic forms \cite{Painl} with $q$ be a function of the modular parameter. Therefore, one can expect some elliptic
parametrization for the ${\rm sl}_N$ case as well. We plan to return to these issues in further publications.

\end{itemize}

}
\end{document}